# A Coordinated Volt/Var Control Scheme for Distribution Systems with High DER Penetration

Yue Shi, *Member, IEEE*, Mesut Baran, *Fellow, IEEE*

*Abstract*—In this paper, a new Volt/Var Control (VVC) scheme is proposed to facilitate the coordination between the conventional VVC devices and the new smart PV inverters to provide an effective voltage control on a system with high PV penetration. The proposed scheme decomposes the problem into two levels. The first level uses Load Tap Changer (LTC) and Voltage Regulators (VRs) to adjust the voltage level on the circuit to keep the voltages along the circuit within the desired range. The second level determines Var support needed from smart inverters to smooth the fast voltage variations while providing effective power factor correction to keep the power losses at minimum. The case study shows that the proposed VVC method is very effective in maintaining acceptable voltages on the system under various operating conditions while meeting the operational constrains. The results also show the computational efficiency of the method.

*Index Terms*—Volt/Var control, coordinated control, smart inverters, DER, distribution.

## I. Introduction

Adoption of Photovoltaics (PV) based systems at both residential and commercial scale has accelerated recently. As more PV systems are integrated into a distribution feeder, they will start affecting the voltage control on the feeder, and when the voltage variations become excessive, some mitigation is needed. Conventional Volt/Var control (VVC) devices on the circuit: Load Tap Changer (LTC), Voltage Regulators (VRs), and capacitor banks (Caps). These devices, are slow acting devices and usually employ local controllers, and hence they cannot provide effective voltage control when PV penetration gets high on the circuit [1]. The variability of PV output can also cause excessive operation of traditional VVC devices [2].

A promising approach recently emerged is the use of the inverters (which serve as the main interface for interconnecting DERs to the grid) to help mitigate the voltage violations [3], [4]. To make effective use of inverters for this purpose, operation of these inverters need to be coordinated with the existing VVC devices. These smart inverters can respond faster to the voltage violations, and local control approaches for these inverters are currently being finalized under IEEE standard 1547 [3]. This paper focuses on coordination of these new smart inverters with the conventional Volt/Var control to improve VVC in systems with considerable PV penetration.

Most of VVC schemes proposed in the literature, aim at determining the set points for the Volt-Var control devices at a given operation point, and thus formulate the VVC as an optimal power flow problem [5]-[11]. However, to prevent excessive VVC device operation due to variability of PV requires that VVC method to consider a large time horizon. Therefore, in this paper we extend the VVC problem and treat it as a tracking problem.

Although the primary goal of VVC is to keep voltages within an acceptable range, the other goals include: 1) Power quality, such as the improvement of voltage profile and power factor. 2) Energy saving, for example power loss reduction and peak demand reduction. 3) Reduction of control cost. A common approach to solve this multi-objective problem is to choose one as the main goal and try to handle the others by adding constraints in the optimization method adopted [6]-[9].

Recently proposed VVC methods usually adopt a model based optimization method [6]-[11]. Since VVC is a complex optimization problem, even for a single operating point, usually some approximation is adopted to solve the problem [12], [13]. Efforts have also been made to develop decentralized schemes [9], [14], [15], rather than SCADA based centralized schemes [6]-[8], in order to minimize the SCADA requirements. However, performance of these methods is usually limited by the lack of system level model [19]-[21]. As a trade-off between centralized and decentralized approach, this paper proposes a master-slave architecture, which facilitates the coordination between the conventional control devices and PV inverters to provide an effective voltage control on a system with high PV penetration. The proposed method adopts a two-level VVC architecture in which smart inverters coordinate with other conventional VVC devices to provide effective voltage control on the feeder and to mitigate excessive operation of LTC and VRs. This approach is computationally efficient and accommodates the operational constraints. Thus, the method can be easily adopted for practical implementation. This is the other contribution of the paper.

The paper is organized as follows: Section II illustrates how the problem is decomposed in two loops, a slower outer loop and a faster inner loop in order to coordinate operation of

This work was supported by ERC Program of the National Science Foundation under Award Number EEC-08212121.

Yue Shi is with North Carolina State University, Raleigh, NC 27606 USA (e-mail: yshi6@ncsu.edu).

Dr. Mesut Baran is with North Carolina State University, Raleigh, NC 27606 USA (e-mail: baran@ncsu.edu).

conventional VVC devices with smart inverters. Section II also presents the methods adopted to solve the sub-problems at each level. Simulation results are presented and discussed in Section III. Finally, Section IV summarizes the advantages of the proposed coordinated VVC scheme and concludes the paper.

## II. COORDINATED VOLT/VAR CONTROL

As indicated above, the goal for VVC is to track both the slower load variations and the faster PV variations on a feeder with the goal of minimizing voltage variations while also avoiding the excessive operation of Voltage Regulators (VRs) and Cap Banks which may occur due to high variability in PV output. To achieve this, we make use of the fact that the conventional VVC devices are intended to operate at a much slower pace than the smart inverters. Also, while the VRs are mainly designed to adjust the voltages, Cap Banks and Smart Inverters are mainly reactive power support devices for power factor correction (and thus helps lowering the power losses). These features of VVC devices helps us to decompose the problem [5]. This paper adopts this approach and decomposes the VVC problem into two loops, a slower voltage control loop and a faster Var compensation loop. Fig. 1 shows the proposed architecture.

*Voltage Control Loop*

The voltage control devices, LTC and VRs, are very effective in adjusting the voltage along the feeder, but they are slow acting devices. Hence, the goal in voltage control loop is to monitor the net load variations along the feeder, filter out the fast variations and have the voltage control devices respond only to slow variations. This prevent excessive device operation. This loop is slow, (which is chosen as 15 min in this paper) and provides the supervisory set points for the LTC and VRs. Details of the method adopted for this module is presented later in this section.

*Var Compensation Loop*

This loop provides supervisory set points for the smart inverters, and since these inverters are fast acting, this loop can be faster than the voltage control loop (we chose it to be 5 minutes). The main goal of this fast acting loop is to respond the over/under voltage violations that may occur due to fast variability in PV output. If there is not enough Var support to correct the voltage violations, then this loop sends a signal to voltage control loop to have the voltage regulation devices to respond immediately. As indicated in Fig. 1, when there is no voltage violation, the control objective changes, and the loop determines Var compensation levels for the inverters in order to minimize the power loss on the feeder. An optimization method is used to determine the proper Var support from the inverters in both cases. Details of this module are given later in this section.

### A. Voltage Control Method

As indicated before, the goal of the voltage control module is to monitor load variations and adjust voltage levels by using the LTC and VRs such that the voltages will remain within the acceptable range within the control window. Hence, as Fig. 1 illustrates, once the module gets a new load update (assumed to be every five minutes), the program first checks if the load change is big enough to cause voltage violation without Var support from inverters. If this is the case, and this condition persists during the control cycle then this indicates that the load has changed enough for the module to adjust voltage levels.

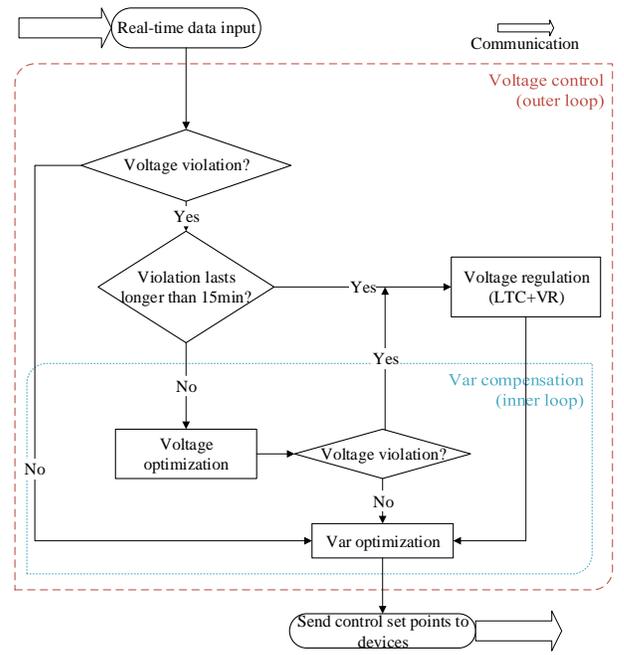

Fig. 1 Two-level coordinated Volt/Var Control

Since, LTC and VRs operate by adjusting their taps, the proper set point can best be determined by calculating the new tap positions needed. Although number of LTC or VRs on a distribution feeder are usually low, the number of possible tap positions each device has (usually 33 tap positions, including the zero-tap position) can make the search space considerably large, as the VRs usually operate on a phase basis. For example, a system with one LTC and one VR has $33 \times 33 \times 33 \times 33 = 1{,}185{,}92$ possible tap position combinations. To reduce the search, a search method based on [16] has been developed. The method is based on the observation that most of the time tap adjustments needed are small. Hence, instead of searching all possible 33 taps, the method considers moving the current tap of each device by two taps up and down: $\Delta Tap = \{0, \pm 1, \pm 2\}$. We found that this search can be reduced further by applying the additional rules given in Fig. 2. As the figure shows, starting from the voltage control device closest to the substation (e.g. LTC), tap position of that device varies while keeping the taps constant for other downstream voltage control devices. Then voltage is calculated for each $Tap_{LTC}(k) \in \Delta Tap$. In Fig. 2, p represents phase a, b, or c. The idea behind the rules is to avoid searching the increasing taps when over-voltage violation occurs and to avoid searching the decreasing taps when under-voltage violation occurs.

Note that there are usually many different tap combinations that will bring the node voltages within the limits. Hence, a criterion is needed to select a solution. In this paper, the tap position combination with least power loss is selected as the criterion, as the power loss reduction is another goal of VVC. A backward-forward distribution power flow solver, DPF, is used to determine how the node voltages change as control variables are varied.

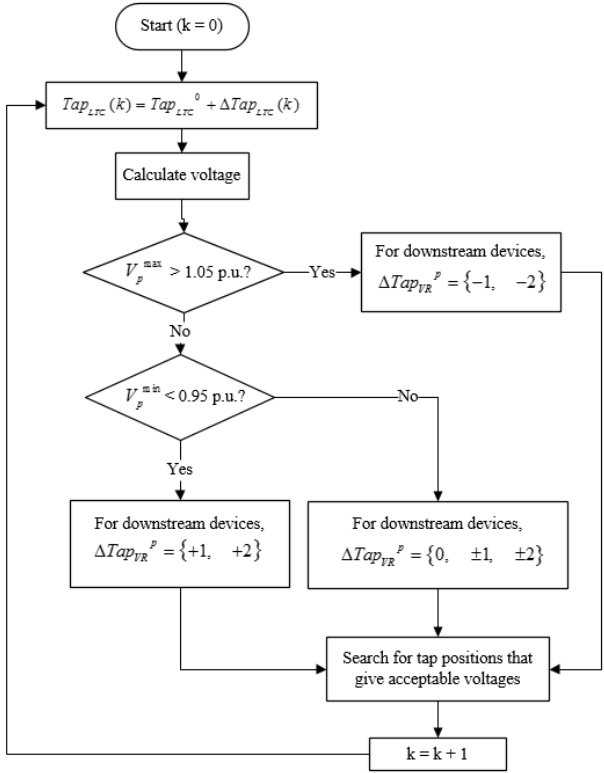

Fig. 2: Search method for Voltage Control loop

*B. Var Compensation Method*

As indicated above, this module uses an optimization method to determine the proper Var support needed from inverters in the system. The objective of Var compensation, under normal conditions, is to improve system efficiency by reducing power loss on the system. However, when there is need for extra Var support for boosting the voltages, the module determines this extra Var support first.

The Var compensation problem is formulated as an optimization problem:

$$\min f(x) \quad (1)$$

$$s.t. \quad g(x,u) = 0 \quad (2)$$

$$V^{min} \leq V \leq V^{max} \quad (3)$$

$$Q^{min} \leq Q_{inv} \leq Q^{max} \quad (4)$$

where $x = [\theta; V]$ are node voltages, and $u$ contains the control variables, i.e. the Var support from smart inverters, $Q_{inv}$ and the equality constraints (2) correspond to power flow equations.

The objective function $f$, as indicated, is the power loss on the system under normal conditions. As the total power loss is the sum of the line losses, $f$ will be:

$$f(x) = P_{loss}(V,\theta) = \sum_{i \neq j} g_{ij} \left[ V_i^2 + V_j^2 - 2V_i V_j \cos(\theta_i - \theta_j) \right] \quad (5)$$

For the second case, when we need to determine the extra Var support, the objective function becomes:

$$f(x) = \sum_{i=1}^{n} \left\{ \omega_i \cdot (V_i - 1.0)^2 \right\} \quad (6)$$

where $\begin{cases} \omega_i = 0 \text{ if } V_i \in [0.95, \ 1.05] \\ \omega_i = 1 \text{ otherwise} \end{cases}$

which aims at determining the minimum Var support needed to bring the node voltages within the limits (which are taken as 0.95 to 1.05 p.u.[17]).

Note that this is a non-linear programming problem with continuous control variables, as inverter Var support is continuously adjustable. Sequential line search methods [12], [13], and Successive Linear Programming (SLP) methods [18] have been proposed to obtain the solution. However, the SLP based method takes much time to update linear model and the linear trust region is usually not a good approximation [19].

In this paper, a gradient based method is adopted to solve the Var problem. Essentially, the proposed gradient based method is a sequential line search method along the direction of steepest descent. The steepest descent (negative gradient) of the objective function with respect to the control variables, $-\nabla f_u$, provides an effective direction for updating the control [20], [21]. This method is also easy to implement as no second derivative is needed. Furthermore, in this application, the past operating point provides a good starting point for the next control update.

Fig. 3 shows how the steepest descent (updated gradient method) is applied to solve the Var problem. As the figure shows, to obtain the optimal solution, the gradient is updated iteratively and an optimal step-size needs to be determined in each steepest descent direction. In this figure, the iterator $i$ is for the outside loop to update the gradient. In each search along the steepest descent, the gradient $\nabla f_u(i)$ is calculated first (details of calculating $\nabla f_u$ is given in the appendix), and then the control variable is updated as

$$u(i) = u(i-1) - \beta^*(i) \cdot \nabla f_u(i) \quad (7)$$

In each gradient update, the challenge is to determine the best step size $\beta^*(i)$ such that the objective function (power loss) is minimized without violating voltage constrains. The best-step size $\beta^*$ can be found by an inner loop search where $\beta$ is updated as

$$\beta(k+1) = \beta(k) \cdot \gamma \quad (8)$$

for some γ > 1. The largest β(k) which achieves the maximum power loss reduction without violating the constraints is selected as $\beta^*$. There're also other methods such as Armijo's rule [12] to determine $\beta^*$. However, adoption of this rule is not easy for a complex function like power loss function [12], [13]. In the proposed method, the initial value β(0) is selected such that the minimum Var support adjustment, $|\Delta u|^{min}$ is some small value, for example 0.01 kVar.

*C. A Master-Slave Control Architecture*

In a smart distribution system, Intelligent Electronic Devices (IEDs) enable communication and processing capability at the feeder level. In this study, each PV smart inverter is assumed to have this capability - and by making use of it- we developed a master-slave based control scheme for VVC rather than conventional centralized SCADA based control architecture which requires all the data from the feeder to be sent to a central control center. In the proposed scheme, the proposed VVC scheme is implemented in a master controller which can sit at a local substation. The master controller then communicates with slave controllers which are implemented IEDs located near to PV clusters. These slave controllers control a group of PV inverters. Details of this approach are given below.

*1) Grouping of the VVC devices*

Since a distribution feeder line section can be rather short, the distance between smart inverters can be short. And therefore, sensitivity of the power loss with respect to $Q_{inv\ i}$ of smart inverters that are close to each other will be very similar. Indeed, when we examine the gradient $\nabla f_u$ we can see that it is the case. This observation indicates that we can group the inverters that are close to each other together and put them under a local "slave controller", so that all the inverters in this group can be controlled together. This is the approach used to group the inverters on a feeder into smaller clusters and form a master-slave architecture.

*2) Master VVC unit and slave VVC units*

To coordinate the control among the groups, a "master controller" is used to act as a supervisory controller and to provide the update signals to slave units based on the proposed method. For Var compensation, the solution $u^*$ is partitioned as shown in Fig. 4 according to the grouping of inverters and then sent to the corresponding slave units. The index $j$ refers to index of VVC slaves, and the vector $u_{mj}^*$ corresponds to all inverters managed by slave $j$. And $p$ is the number of groups.

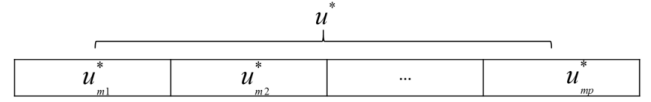

Fig. 4 Partition of the solution of Var optimization

After receiving $u_{mj}^*$, slave $j$ allocates the reactive power injection based on the capacity of reactive power support of inverters. For each smart inverter $i$ under VVC slave $j$:

$$u_{i,j} = \left(\sum_{i=1}^{N_j} u_{i,j}^*\right) \cdot \frac{Q_{i,j}^{\max}}{\sum_{i=1}^{N_j} Q_{i,j}^{\max}} \quad (9)$$

where $N_j$ is the total number of inverters under the slave $j$, $Q_{i,j}^{\max}$ is the reactive power capacity of inverter $i$ under slave $j$.

Note that the inverter Var capacity depends on its power output. To provide sufficient reactive power support when PV generation is high, the inverters are oversized with 25% more than peak generation, i.e. $S_N = 1.25 \times P_{\max}^{PV}$. Therefore, the available reactive power varies as the PV output, $P_{PV}$, changes:

$$Q^{\max}(t) = \sqrt{S_N^2 - P_{PV}(t)^2} \quad (10)$$

### III. TEST RESULTS

*A. Test System*

To demonstrate the effectiveness of the proposed VVC scheme, the IEEE 34 node test feeder [22] is modified with PVs connected to all nodes with load. Since these inverters can provide sufficient reactive power to improve the voltage profile, only one VR is needed in this modified system, and hence VR#2 and Caps are removed. A three-phase controlled LTC is placed at the substation, and three phases of the VR is controlled independently. Fig. 6 shows how the inverters are grouped based on sensitivity/gradient analysis. As the figure shows, in this case 20 inverters are divided into 6 groups. To simulate operation of the test feeder on a typical day, a 24-hour PV and load profile with a 5-min resolution, shown in Fig. 7, is used. Note that this corresponds to a challenging case as there is high PV variability causing even reverse power flow during short periods on the system.

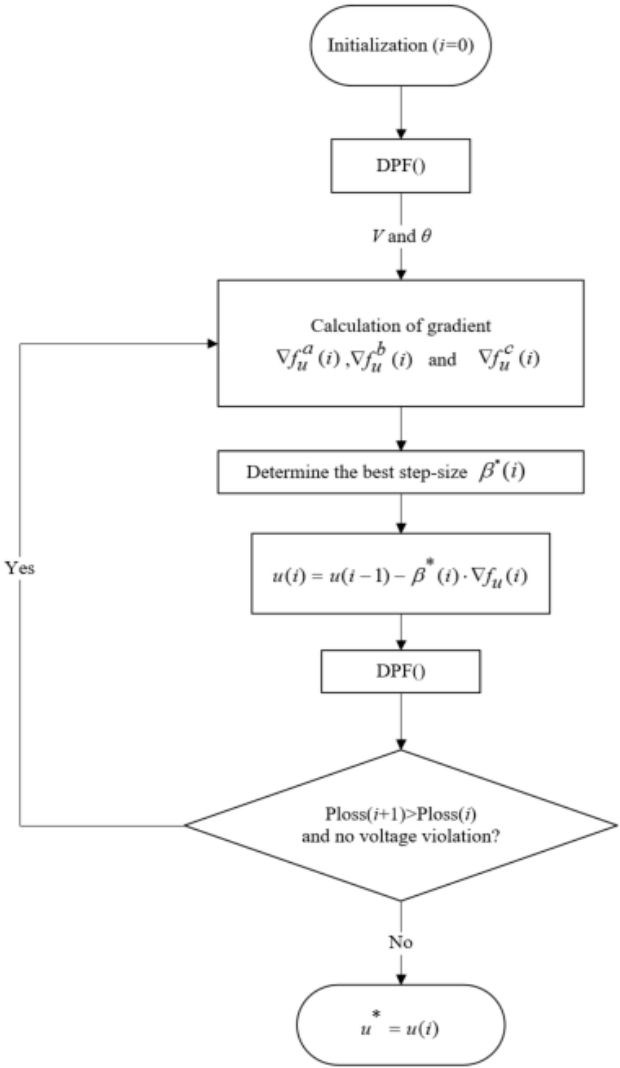

Fig. 3 Updated gradient based Var optimization

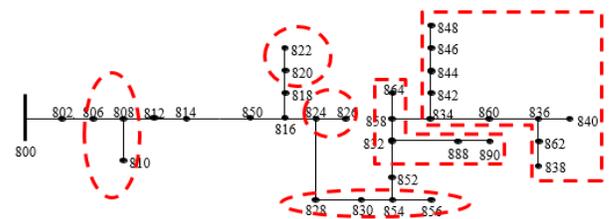

Fig. 6  Grouping of the smart inverters on the IEEE 34 nodes system

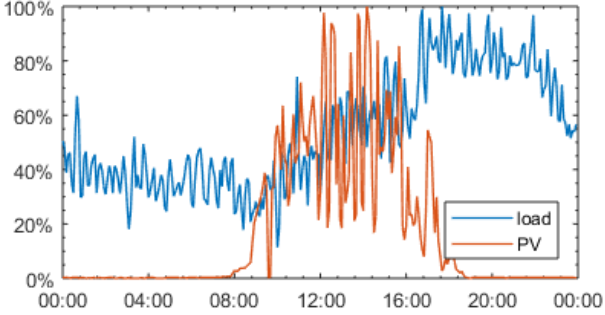

Fig. 7 A 24-hour PV and load profile

### B. Performance of Var compensation method

To validate the performance of the proposed gradient based method for Var compensation, two operating conditions are simulated.

*Case 1: heavy load*

This case illustrates the performance of the proposed Var compensation method during the peak load condition. At the initial starting point, there is no voltage violation, and hence the Var compensation tries to minimize the total power loss on the feeder. Initial Var compensation for the inverters is chosen such that it compensates for the reactive power of the local load.

The initial step-size, $\beta(0)$, corresponds to $|\Delta u|^{min} = 0.02\ kVar$ and $\gamma$ is selected to be 1.1.

Fig. 8 shows the power loss reduction obtained during the iterations of the gradient based method. Fig. 9 shows typical convergence profile of the steepest descent method: good monotone convergence during the first few iterations and then convergence slows down as the solution gets close to the optimal point. This indicates that in practice only a few iterations are needed to get a good solution. Furthermore, our investigations indicate that updating the gradient at every iteration does not improve the objective value considerably, and thus the gradient update may not be needed. Indeed, the solution given in Fig. 8 does not update the gradient.

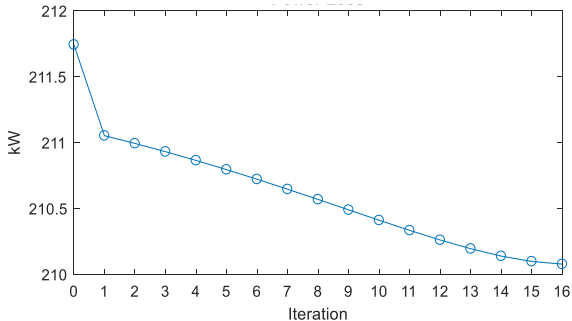

Fig. 8 Power loss of gradient based of Var optimization

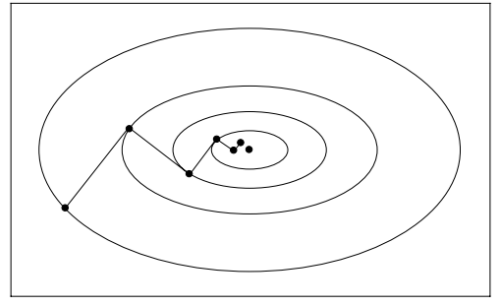

Fig. 9 Convergence profile of steepest descent method [12]

To verify the effectiveness of the proposed gradient method, the sequential linear programing (SLP) method is applied to solve the same Var optimization problem. Table I compares the solution (Var compensation) from the two methods. As Table 1 shows, the proposed gradient method reduces the power loss to 210.077 kW which is close to the result of SLP method, 209.992 kW. Fig. 10 shows the new voltage profiles obtained by two methods. Note that the new voltage profiles of two methods are nearly the same, and hence, these results confirm that proposed method provides very good solution. Finally, note that the proposed method is computationally much more efficient; the runtime for gradient method is 1.002s which is much shorter than SLP's runtime, 5.902s.

TABLE I. RESULTS OF GRADIENT VVC V.S. LP VVC

|  | *Iterations* | *Vmin (p.u.)* | *Vmax (p.u.)* | *Loss(kW)* | *Loss Reduction(kW)* |
|---|---|---|---|---|---|
| Before VVC |  | 0.965 | 1.05 | 211.742 |  |
| After Gradient VVC | 16 | 0.972 | 1.05 | 210.077 | 1.665 |
| After LP VVC | 21 | 0.973 | 1.05 | 209.992 | 1.750 |

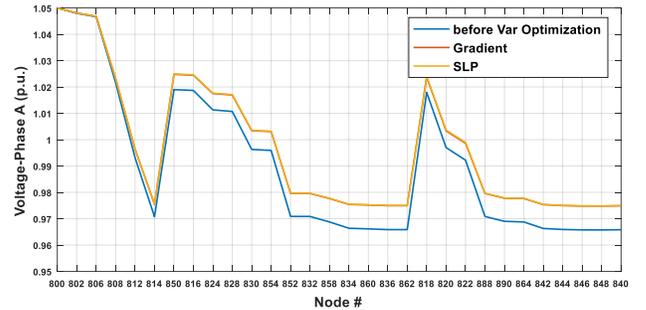

Fig. 10 Phase A feeder voltages for Case 1

*Case2: Light load*

Case 2 simulates the extra Var support needed during a light load condition where voltage violation occurs due to sudden change in PV output (event corresponds to 10:05 am in Fig. 7). In this case, LTC and VR do not respond to this event, but the Var support module reacts and determines extra Var support needed from the inverter to eliminate the voltage violation. Here, the gradient is updated because as the indicator variables $w_i$ change, they affect the gradient vector considerably. Fig. 11 shows how the node voltages during the iterations of the method. As the figure shows, the node voltages move towards the acceptable ranges quite rapidly and it takes only a few iterations to bring the voltages within 0.95 – 1.05 p.u.

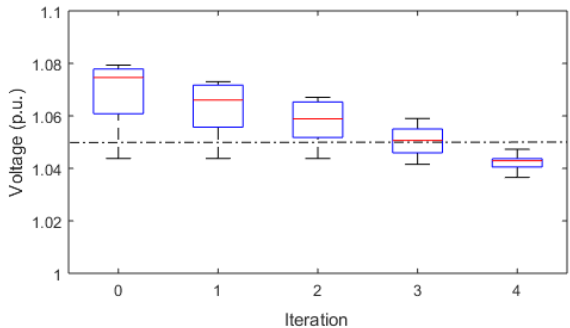

Fig. 11 Box plot of voltage variation in voltage optimization

*C. Performance of the VVC method*

A 24-h simulation is performed to verify the effectiveness of the proposed scheme. At the beginning of every control period of 5 minute, data for the new operating point (net load at each node) is obtained first by running the DPF. The VVC method is then run to determine the updated settings for the VVC devices. These new settings are then sent to the system (which is simulated using DPF) and this control cycle is repeated in 5 minute increments.

Fig. 12 shows how the voltage control devices, LTC and VR, are controlled. From the results we can see that there is no excessive operation of the devices. Fig. 13 illustrates the reason for each operation. Markers " * " indicates the LTC/VR operation under normal conditions, i.e. the net load change in 15-min period is large enough to warrant adjustment of tap settings. Markers " o " indicate the tap adjustment needed when there is not enough Var support from inverters to keep the voltages within limit, therefore these operations can mainly be considered as the extra operations due to power variability caused by PVs. Markers " + " on the figure indicate Var support from inverters when voltage violation occurs due to sudden change in PV output. These points thus show that indeed Var support helps to respond the sudden voltage violations quickly and this in turn reduces the excessive operation of voltage regulation devices. Fig.14 confirms that the system maximum and minimum voltages are always kept within the range 0.95 – 1.05 p.u.

Fig. 15 shows the Var support from the inverter at Node 830 phase B. The simulation indicates that Var support is adjusted throughout the day for both loss reduction as well as for voltage control, and majority of the smaller adjustments are for loss reduction under normal conditions (ie. no extra var support for voltage control).

Fig. 16 shows the power loss profile under both the proposed method and the conventional VVC (used on the IEEE 34 node test feeder) during the day. The results indicate that the proposed VVC provides considerable power loss reduction compared to the conventional VVC scheme: the total energy loss during the day in this case is 36.7% lower than that of the conventional VVC scheme. Fig. 17 shows the Phase A tap operations of two VRs in the original system. As expected, with variability of system PV output and load, as shown in Fig. 7, conventional VR control results in excessive tap operations.

To verify the effectiveness of the proposed voltage control scheme, the 24-h simulation is repeated with a different voltage control scheme. In this case, an exhaustive search method employed for determining the optimal tap settings for voltage control devices when they need to respond to slow load

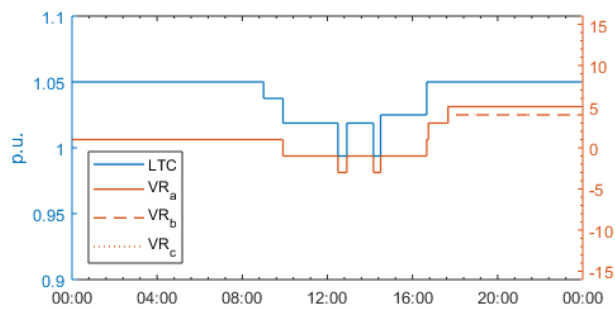

Fig. 12 LTC and VR tap positions for 24h

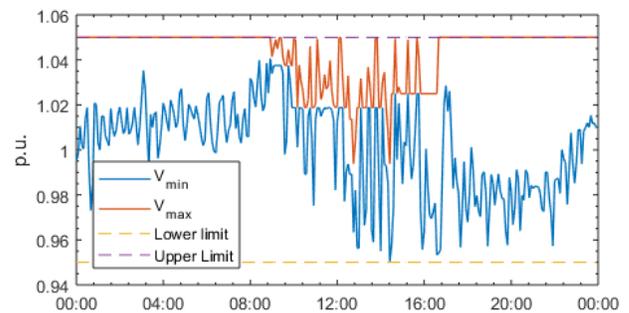

Fig. 14 System max/min feeder voltage profile

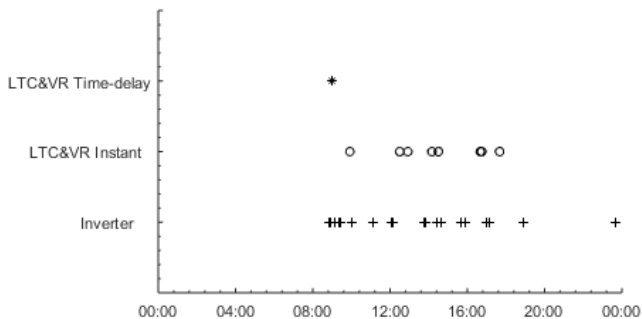

Fig. 13 Operation sequence of VVC devices

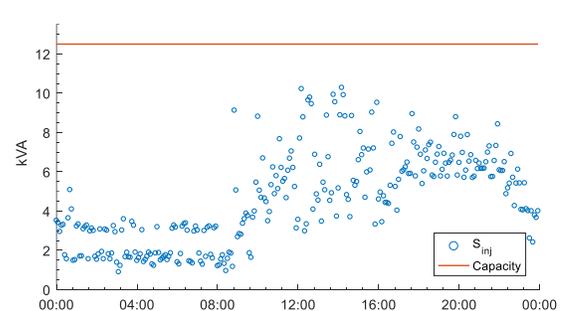

Fig. 15 Apparent power at inverter 830 phase B

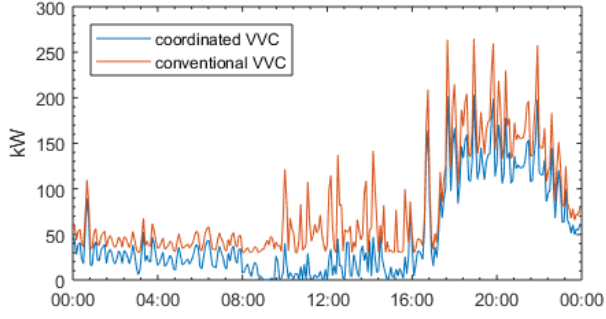

Fig. 16  System power loss profile

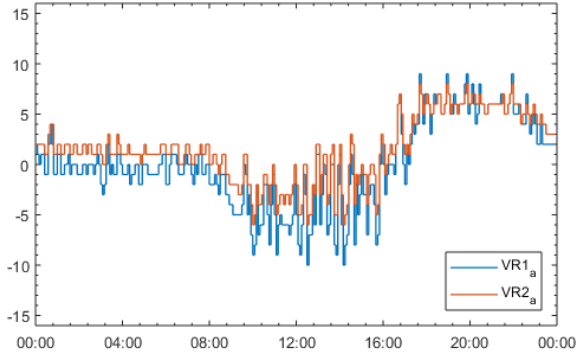

Fig. 17  VR tap operation profile under conventional VVC

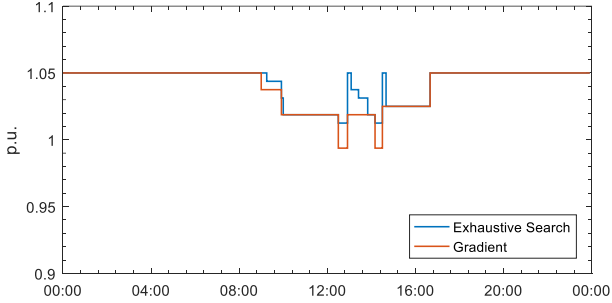

Fig. 18  LTC tap position profile

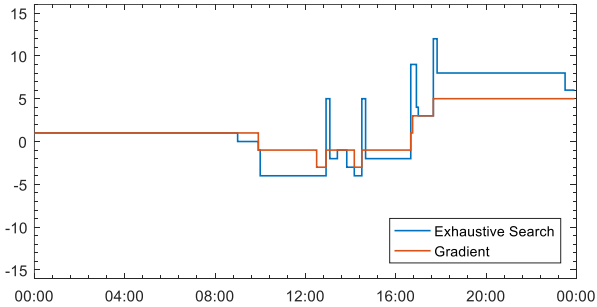

Fig. 19  VR tap position profile

changes. Fig. 18 and Fig. 19 show the LTC and VR Phase A tap profiles for the two methods. The two figures show that the exhaustive search method moves the taps more often with finer adjustments, as expected. The main benefits of these adjustments are better loss reduction, as shown in Fig. 20. However, the cost associated with excessive tap operation may outweigh the additional improvement in power loss reduction.

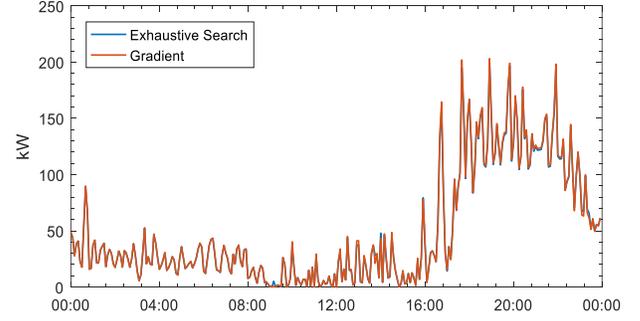

Fig. 20  System power loss profile

Finally, the computation time of the proposed coordinated VVC method is quite low, 288 executions of the program (during 24-h simulation) takes 123s (0.43s per operating point on average). This clearly indicates the feasibility of the method in practice.

## IV. Conclusions

In this paper, a two-level coordinated Volt/Var control is proposed. The IEEE 34 node test feeder is used to test the performance of proposed VVC scheme. The results show that the method handles the three operational requirements for VVC effectively: maintaining the node voltages within limits while avoiding excessive device operations, and reducing power losses. The test results also show that proposed gradient method for Var compensation is very effective in determining the Var support from DERs. Results based on a daily operation also verify that the method minimizes the operation of the LTC and VR by using Var support from smart inverters to respond to the voltage variations caused by variable PV output.

Runtime results also show that the proposed VVC method is fast and therefore appropriate for practical implementation.

## V. Appendix

*Calculation of Gradient* $\nabla f_u = \dfrac{\partial f}{\partial u}$

Since the objective function in the Var compensation problem in Section II is function of x only, we can use chain rule to calculate the reduced gradient $\nabla f_u$ as follows:

$$\frac{\partial f}{\partial u} = \frac{\partial f}{\partial g} \cdot \frac{\partial g}{\partial u} \tag{11}$$

$$\frac{\partial f}{\partial x} = \frac{\partial f}{\partial g} \cdot \frac{\partial g}{\partial x} \tag{12}$$

here $\dfrac{\partial f}{\partial u}$ and $\dfrac{\partial f}{\partial x}$ are vectors. Equation (12) can be rearranges as

$$\frac{\partial f}{\partial g} = \frac{\partial f}{\partial x} \cdot \left[\frac{\partial g}{\partial x}\right]^{-1} \tag{13}$$

Note that

$$\frac{\partial g}{\partial x} = \begin{bmatrix} \dfrac{\partial \Delta P}{\partial \theta} & \dfrac{\partial \Delta P}{\partial V} \\ \dfrac{\partial \Delta Q}{\partial \theta} & \dfrac{\partial \Delta Q}{\partial V} \end{bmatrix} = J$$

is the Jacobian matrix of power flow equations.

Substituting $\dfrac{\partial f}{\partial g}$ in (11) by (13), the reduced gradient can be calculated as:

$$\nabla f_u = \frac{\partial f}{\partial u} = \left( \frac{\partial f}{\partial x} \cdot \left[ \frac{\partial g}{\partial x} \right]^{-1} \right) \cdot \frac{\partial g}{\partial u} \qquad (14)$$

Note that $\dfrac{\partial g}{\partial u}$ can be derived from (2):

$$\frac{\partial g}{\partial u} = \begin{bmatrix} \dfrac{\partial \Delta P}{\partial Q_{inv}} \\ \dfrac{\partial \Delta Q}{\partial Q_{inv}} \end{bmatrix} = \begin{bmatrix} 0 \\ I \end{bmatrix} \qquad (15)$$

And $\dfrac{\partial f}{\partial x}$ can be obtained directly:

$$\frac{\partial f}{\partial x} = \begin{bmatrix} \dfrac{\partial f}{\partial \theta} \\ \dfrac{\partial f}{\partial V} \end{bmatrix} \qquad (16)$$

$$(17)$$